\documentclass[reprint,showpacs,preprintnumbers,amsmath,amssymb,prl]{revtex4-1}
\usepackage{amsmath}
\usepackage{amsfonts}
\usepackage{amsthm}
\usepackage{graphics}
\usepackage{graphicx}
\usepackage{subfigure}
\usepackage{amssymb}
\usepackage {mathrsfs}
\usepackage{braket}

\usepackage{color}
\usepackage{url}
\usepackage[abs]{overpic}
\usepackage[usenames,dvipsnames]{xcolor}

\def\v{\textbf{v}}

\begin{document}

\title{Quantum walk hydrodynamics}

\author{Mohamed Hatifi}
\email{hatifi.mohamed@gmail.com}
\affiliation{Aix Marseille Universit\'{e}, CNRS, \'{E}cole Centrale de Marseille, Institut Fresnel UMR 7249, 13013 Marseille, France}
\author{Giuseppe Di Molfetta}
\email{giuseppe.dimolfetta@lif.univ-mrs.fr}
\affiliation{Aix-Marseille Univ., CNRS, LIF, Marseille, France and Departamento de F{\'i}sica Te{\'o}rica and IFIC, Universidad de Valencia-CSIC,
Dr. Moliner 50, 46100-Burjassot, Spain}

\author{Fabrice Debbasch}
\email{fabrice.debbasch@gmail.com}
\affiliation{LERMA, UMR 8112, UPMC and Observatoire de Paris, 61 Avenue de l'Observatoire  75014 Paris, France}
\author{Marc Brachet} 
\email{brachet@physique.ens.fr}
\affiliation{Laboratoire de Physique Statistique, 
\'{E}cole Normale Sup{\'e}rieure, 
PSL Research University;
UPMC Univ Paris 06, Sorbonne Universit\'{e}s;
Universit\'{e} Paris Diderot, Sorbonne Paris-Cit\'{e};
CNRS;
24 Rue Lhomond, 75005 Paris, France
}
\date{\today}
\begin{abstract}
A simple Discrete-Time Quantum Walk on the line is revisited and given an hydrodynamic interpretation through a novel relativistic generalization of the Madelung transform. Numerical results are presented which show that suitable initial conditions indeed produce hydrodynamical shocks. An analytical computation of the asymptotic quantum shock structure is presented. The non-relativistic limit is explored in the Supplementary Material (SM).
\end{abstract}
\pacs{03.67.-a, 47.37.+q, 47.40.-x, 67.10.-j}
\keywords{gggggg}
\maketitle

\paragraph{Introduction}Quantum walks (DTQWs) are unitary quantum automata that can be viewed as formal generalizations of classical random walks. Following the seminal work of Feynman \cite{FeynHibbs65a} and Aharonov \cite{ADZ93a} they were considered in a systematic way by Meyer \cite{Meyer96a}. DTQWs have been realized experimentally with a wide range of physical objects and setups \cite{Schmitz09a, Zahring10a, Schreiber10a, Karski09a, Sansoni11a, Sanders03a, Perets08a}, and are studied in a large variety of contexts, ranging from quantum optics \cite{Perets08a}
to quantum algorithmics \cite{Amb07a, MNRS07a}, solid-state physics \cite{Aslangul05a, Bose03a, Burg06a, Bose07a} and biophysics \cite{Collini10a, Engel07a}.
The aim of this Letter is to show, through both literal  and numerical computations, that QWs can also be used to model quantum fluid dynamics (QFD). 

We focus on a simple spatially homogeneous and time independent DTQW on the line whose continuous limit is identical to the free Dirac equation in flat 2D space-time. We then introduce a new relativistic generalization of the Madelung transform which maps this Dirac equation into a 2D dispersive hydrodynamics for relativistic quantum fluids. 
In the non relativistic limit, the two component spinor which obeys the Dirac equations degenerates into a single wave-function which obeys the Schr\"odinger equation, which can also be viewed as the continuous space limit of a continuous time quantum walk. The relativistic Madelung transform then becomes the usual Galilean Madelung transform. To prove that the hydrodynamical vision goes beyond a mere rewriting of the equations, we demonstrate through direct numerical simulations that the DTQW, with suitable initial conditions, actually models QFD shocks. We also present an analytical computation of the asymptotic Galilean shock structure through Pearcey integral commonly used in Optics.
 
\paragraph{The DTQW}
The Hilbert space of the DTQW is the tensor product
$\mathscr{H}_{p}\otimes\mathscr{H}_{s}$, where $\mathscr{H}_{p}$ is the discrete line with basis ${\ket{n}}$, $n \in \mathbb{Z}$ and $\mathscr{H}_{s}$ is the `spin'-space with basis vectors
$\ket{L} = (1, 0)^T$ and $\ket{R} = (0, 1)^T$.
The evolution is controlled by the unitary operator
$U = TC$, where 
$T=\sum\limits_{n}[\ket{n-1,L}\bra{n,L} + \ket{n+1,R}\bra{n,R}]$
is the translation operator
and $C =  e^{-i\theta\sigma_{1}}$ is the quantum coin operator defined from the first Pauli matrix $\sigma_1$ and an arbitrary constant angle $\theta$.
The explicit evolution equation of the walk reads:
\begin{equation}
 \left[ \begin{array}{c} \Psi_{L}(j+1, n-1)\\ \Psi_{R}(j+1, n+1) \end{array} \right]=\left[ \begin{array}{cc} \cos \theta & -i \sin \theta \\
-i \sin \theta & \cos \theta \end{array} \right]\left[\begin{array}{c} \Psi_{L}(j,n) \\ \Psi_{R}(j,n) \end{array} \right]
\label{eq:DTQW}
\end{equation}
where the index $j$ represents the iteration or discrete time. 

\paragraph{Continuous Limit} Introduce now two positive real numbers $m$ and $\epsilon$, choose $\theta(\epsilon, m) = \epsilon m$, consider that $\Psi_{L/R}(j, n)$ are the values taken by some
differentiable functions $\Psi_{L/R}(t, x)$ at point $t_j = j \epsilon$ and $x_n = n \epsilon$. Equation ({\ref{eq:DTQW}) then admits a continuous limit which 
coincides with the Dirac equation \cite{DMD14}
\begin{equation}
i\gamma^{\mu}\partial_{\mu} \psi-m\psi=0 \, ,
\label{eq:Dirac2}
\end{equation}
where $\psi = (\Psi_L, \Psi_R)^T$, $\gamma^0 = \sigma_1$, $\gamma^1 = i \sigma_2$ 
($\sigma_2$ is the second Pauli matrix) and
$\hbar = c = 1$. The mass $m$ is thus homogeneous to the inverse of a length.

Dirac equation 
\eqref{eq:Dirac2} 
can be obtained from the Lagrangian density
$\mathcal{L}=\frac{i}{2}\left(\overline{\psi}\gamma^{\mu}\partial_{\mu}\psi-\partial_{\mu}\overline{\psi}\gamma^{\mu}\psi\right)-m\overline{\psi}\psi$
where $\overline{\psi}=\psi^{\dagger}\gamma^{0}$.
The associated particle current is $j^{\mu}=\overline{\psi}\gamma^{\mu}\psi$ and the stress energy tensor reads
$T^{\mu\nu}=\frac{i}{4}[\overline{\psi}\gamma^{\mu}\partial^{\nu}\psi-\partial^{\nu}\overline{\psi}\gamma^{\mu}\psi+(\mu\leftrightarrow\nu )]$.
Both $j$ and $T$ are conserved {\sl i.e.} $\partial_{\mu} T^{\mu\nu}=0$ and $\partial_{\mu} j^{\mu}=0$. Note that the above Lagrangian density leads to a symmetric canonical 
stress-energy tensor. 

\paragraph{New variables}
The definition of $j$ leads to $j^{0} = |\psi_{R}|^{2}+|\psi_{L}|^{2}$ and $j^{1} = |\psi_{R}|^{2}-|\psi_{L}|^{2}$.
Note that $(j^0)^2 - (j^1)^2 = 4 |\psi_{L}|^{2}  |\psi_{R}|^{2}\ge 0$ so that the current $j$ is necessarily timelike or null.
We then introduce $\varphi_\pm = \varphi_{L} \pm \varphi_{R}$
where $\varphi_{L/R}$ is the phase of $\Psi_{L/R}$
and replace the variables $(\rho_L, \rho_R, \phi_L, \phi_R)$ by $(j^0, j^1, \varphi_+, \varphi_-)$. 
In particular, the spinor $\psi$ now reads
\begin{equation}
\psi ({\bf x},t)=\frac{1}{\sqrt{2}} e^{i \varphi_+/2}\left[\begin{array}{c} \sqrt{j^{0} - j^{1}}e^{i\varphi_{-}/2} \\ \sqrt{j^{0} + j^{1}}e^{-i\varphi_{-}/2} \end{array} \right]
 \label{eq:psi}
\end{equation}
and $\varphi_+/2$ can be viewed as the global phase of 
$\psi$.

In terms if these new variables, the Lagrangian density  
and the stress energy tensor 
read
$\mathcal{L}=-m(j_\mu j^\mu)^{1/2}\cos\varphi_{-}-\frac{1}{2}\left(j^{\mu}\partial_{\mu}\varphi_{+}-\epsilon^{\mu\nu}j_{\nu}\partial_{\mu}\varphi_{-}\right) $
and
$T^{\mu\nu}=-\frac{1}{4}\left( j^{\mu}\partial^{\nu}\varphi_{+}-\epsilon^{\mu\alpha}j_{\alpha}\partial^{\nu}\varphi_{-}+(\mu\leftrightarrow\nu )\right)$,
where $\epsilon^{\mu\nu}$ denotes the completely antisymmetric symbol of rank two, with the convention $\epsilon^{01}=-\epsilon^{10} = 1$.

The dynamical equations derived from ${\mathcal L}(j^0, j^1, \varphi_+, \varphi_-)$ are
\begin{eqnarray}
\epsilon^{\mu}{}_{\alpha}\partial_{\mu}j^{\alpha}&=&2m (j_\mu j^\mu)^{1/2} \sin\varphi_{-} \label{eq:madirac1}\\
m\cos\varphi_{-}\, j^{\mu}&=&-\frac{1}{2}\,  (j_\mu j^\mu)^{1/2}\, (\partial^{\mu}\varphi_{+}+\epsilon^{\mu\nu}\partial_{\nu}\varphi_{-}) \label{eq:madirac}\\
\partial_\mu j^\mu &=& 0 \label{eq:madiraccont}.
\end{eqnarray} 

\paragraph{Dirac quantum hydrodynamics}
Since $j$ is time-like or null, one can define the density $n$ of the $(1 + 1)$D Dirac fluid by $n = (j_\mu j^\mu)^{1/2}$.
We now suppose that $j$ is not null and define the vector $u = j/n$ as the 2-velocity of the fluid, normed to unity. 
The two variables $j^0$ and $j^1$ can then be replaced by $n$ and $u^1$ {\sl i.e.} the density and the spatial part of the fluid 2-velocity. 
Equation (\ref{eq:madirac}) can then be re-written as
$m\cos\varphi_{-}\, u^{\mu}=-\frac{1}{2}\, (\partial^{\mu}\varphi_{+}+\epsilon^{\mu\nu}\partial_{\nu}\varphi_{-})$
and, in this form, brings to mind the standard relation $\frac{w}{n}u^{\mu} =-\partial^{\mu}\varphi$ which links the velocity $u$ of a relativistic potential flow to its potential $\varphi$, the enthalpy per unit volume $w$ and 
the particle density $n$.
We thus retain $w = mn\cos\varphi_{-}$ as the enthalpy per unit volume of the $(1 + 1)$D Dirac fluid.
The velocity field $u$ then derives from two potentials. One is $\varphi_+ / 2$ {\sl i.e.} 
the global phase of the spinor $\psi$ and contributes to $u$ in the standard way. The other potential is the phase differential $\varphi_-/2$ and contributes to $u$ in a non-standard way, by contraction of its gradient with the $(1 + 1)$D completely antisymmetric symbol.

Using (\ref{eq:madirac}), one then finds that 
\begin{equation}
T^{\mu\nu}= w u^{\mu}u^{\nu}+\frac{n}{2}\left( \epsilon^{\mu\alpha}u_{\alpha}\partial^{\nu}\varphi_{-}+u^{\mu}\epsilon^{\nu\alpha}\partial_{\alpha}\varphi_{-}\right), \label{eq:tuu}
\end{equation}
to be compared with the stress-energy tensor $T^{\mu\nu}=wu^{\mu}u^{\nu}-p\eta^{\mu\nu}$ of a relativistic perfect fluid of pressure $p$.
The pressure of the Dirac fluid thus vanishes. This is not surprising because pressure in spin 0 superfluids is generated by interaction terms and there is no interaction in the free Dirac equation 
derived above. The last two terms on the right-hand side of  \eqref{eq:tuu} depend on the gradient of $\varphi_-$ and, thus, 
on the gradient of $w/n$. Indeed, the definition of $w$ leads to $\sin^2 \varphi_- = 1 - (\frac{w}{m n})^2$ and $\sin \varphi_- d\varphi_- = - d (\frac{w}{m n})$,
so that, if $w \ne nm$, 
\begin{equation}
\partial_\mu \varphi_- = - \sigma \frac{\partial_\mu (\frac{w}{m n})}{\left(1 - (\frac{w}{m n})^2\right)^{1/2}}
\end{equation}
where $\sigma$ is the sign of $\sin \phi_-$. As for relativistic spin 0 superfluids, the two extra-terms in the above expression of the stress-energy tensor thus depend on the gradient of a thermodynamic function (the enthalpy per particle $w/n$) and are therefore best viewed as generalized `quantum pressure' terms. 
As shown in the SM the two component spinor which obeys Dirac equation degenerates, in the Galilean limit, into a single wave-function which obeys the Schr\"odinger equation and the relativistic hydrodynamics degenerates into the usual Madelung hydrodynamics. 

\paragraph{Numerical shock simulation}
The above generalization of the Madelung transform strongly suggests that the original DTQW can be used to simulate quantum flows. 
First note that a general positive energy plane wave solution of \eqref{eq:Dirac2} can be written as 
(see \eqref{eq:psi}-\eqref{eq:madiraccont})
$j^0=n \sqrt{1+q^2}$, $j^1=n q$, $\varphi_+/2=-m \left(\sqrt{1+q^2} t - q x\right)$, $\varphi_-=0$,
where $q$ denotes both wave-number and momentum in unit of $m$ (remember $\hbar = c = 1$).
The spinor $\Psi_L=\sqrt{\sqrt{1+q^2}-q} e^{i m \phi}/\sqrt{2}$,  $\Psi_R=\sqrt{\sqrt{1+q^2}+q}e^{i m \phi}/\sqrt{2}$ thus describes, at $t=0$,
a unit density fluid ($n=1$) in motion with constant velocity  $u^1$ given by 
$u^1= q= \partial \phi/\partial x$.
In order to simulate quantum flows, we now select
the initial data
\begin{equation}
\phi=\frac{q_{\rm max}}{m}\left[{\cos (x)+\frac{1}{3} \cos (3 x)+\frac{1}{2} \cos (2 x+0.9)}\right],
\label{eq:indat}
\end{equation}
which, with $q_{\rm max}=m u_{\rm max}$ corresponds to the velocity field $u^1=u_{\rm max}( -\sin (x)-\sin (3 x)- \sin (2 x+0.9))$. 
The evolution of this initial condition through the DTQW for various values of $m$ and constant $q_{\rm max}$ (the larger the mass, the less relativistic the propagation) is displayed in Fig.\ref{fig:EiffelTower}. 

Note that a similar (but somewhat simpler) type of initial condition 
$\phi=q_{\rm max}\cos (x)/m$ 
has already been used in the cosmological context to simulate the dynamics of (i) a non-quantum cosmological fluid through the non-linear Schr\"odinger equation \cite{Coles2003}  (ii) a Bose-Einstein condensates of axions \cite{Sikivie2009}. Fig.\ref{fig:EiffelAndSimpleShock} (compare Figs. \ref{fig:EiffelAndSimpleShock}.a and \ref{fig:EiffelAndSimpleShock}.b) shows that this simpler initial data generates a single symmetric shock.

\begin{figure}
\includegraphics[width=4.2 cm]{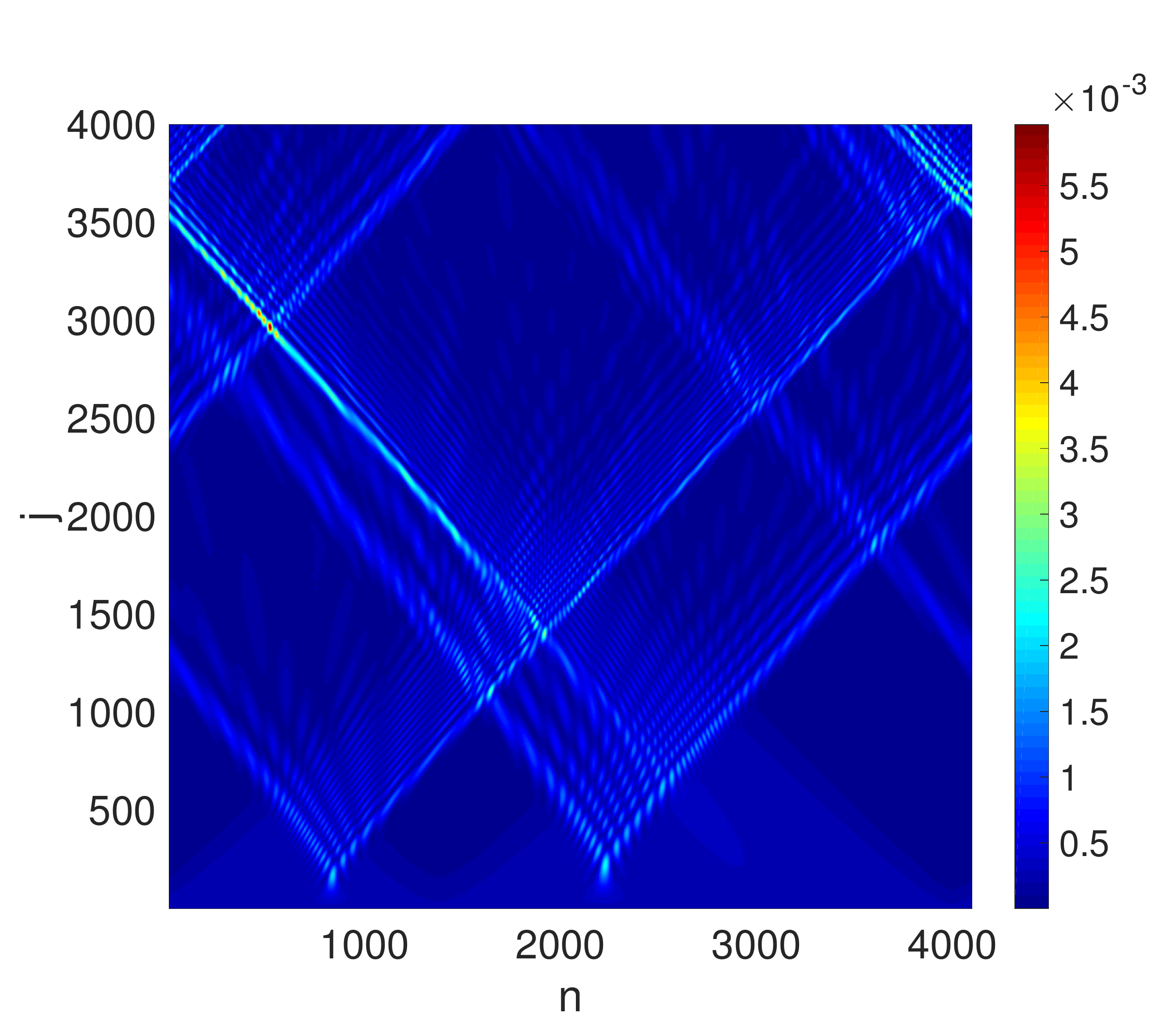}
\includegraphics[width=4.2 cm]{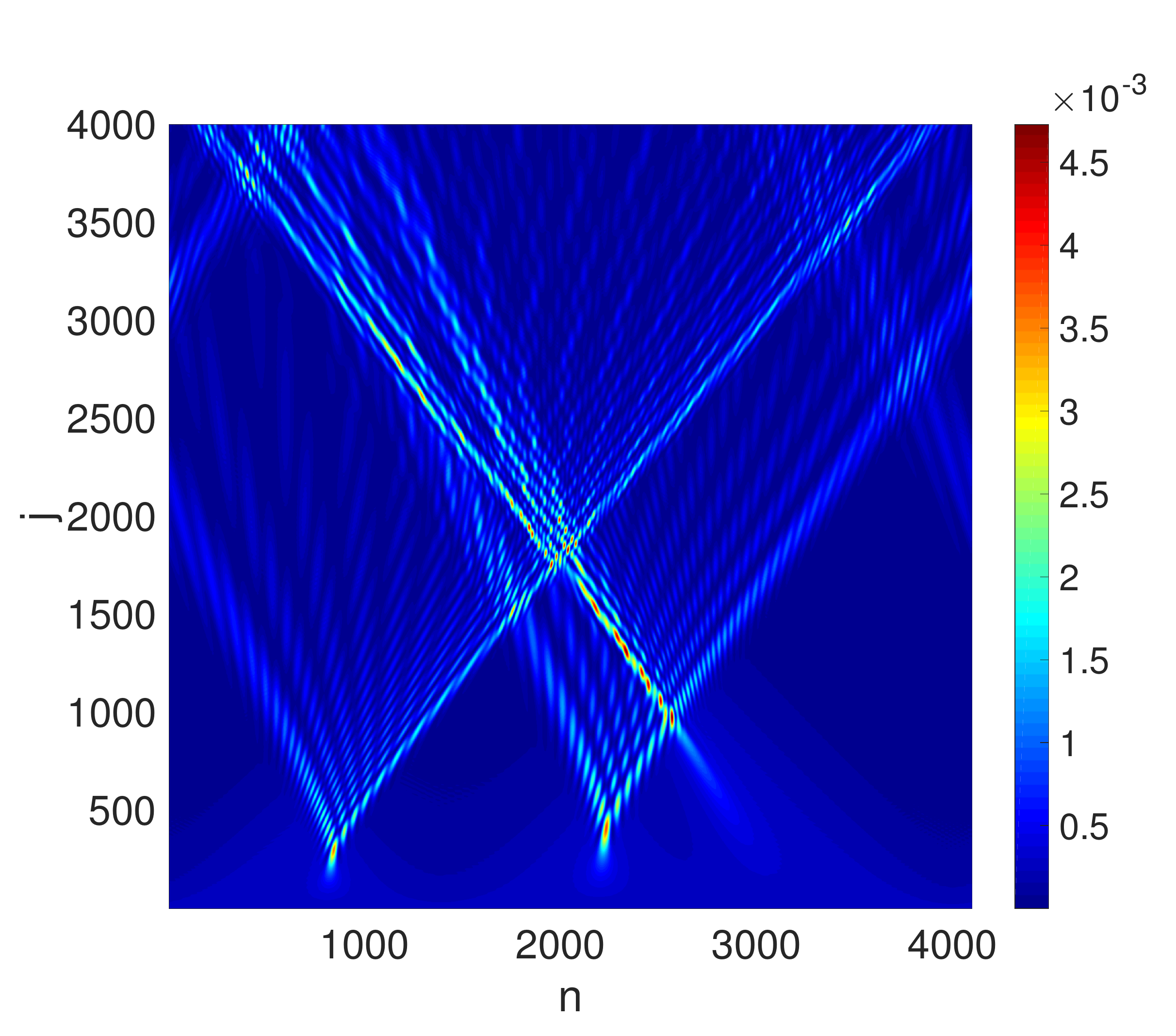}
\includegraphics[width=4.2 cm]{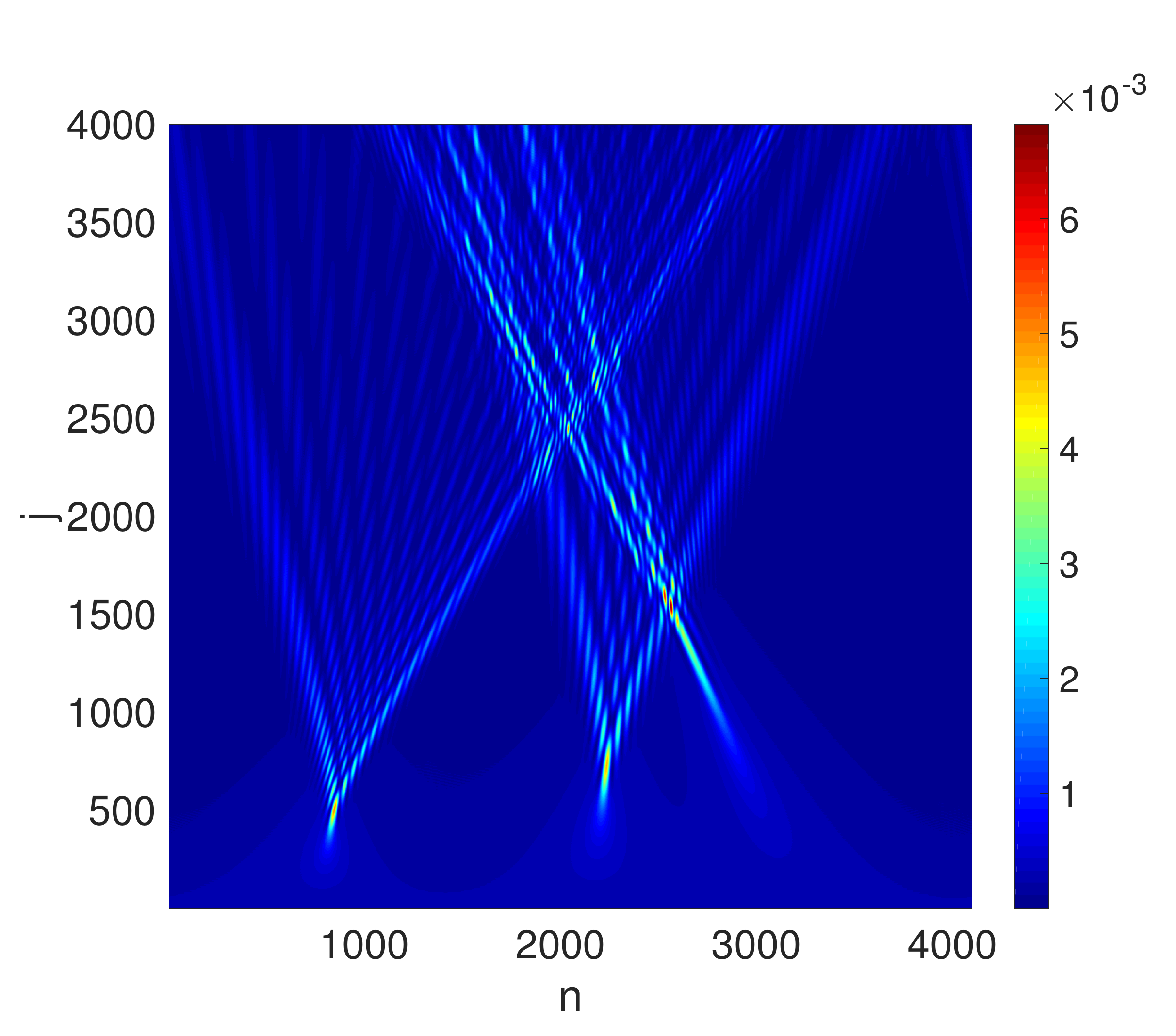}
\includegraphics[width=4.2 cm]{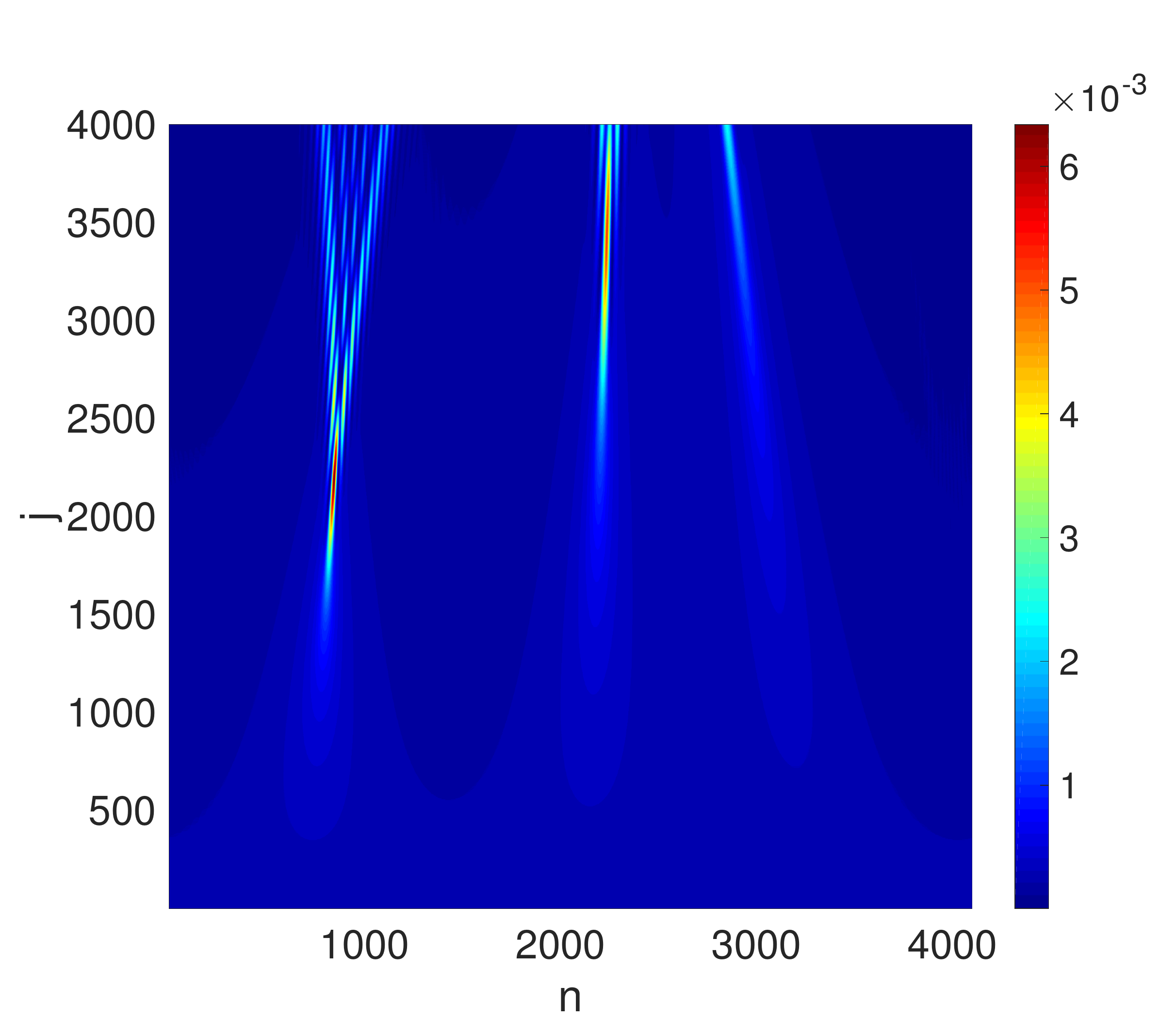}
\caption{(color online) Unitary evolution of density $j^0=|\Psi_L|^2+|\Psi_R|^2$ for the DTQW defined in Eq.\eqref{eq:DTQW} with initial data: $\phi=q_{\rm max}\left[{\cos (x)+\frac{1}{3} \cos (3 x)+\frac{1}{2} \cos (2 x+0.9)}\right]/m$ where $m:=$ $25.6$, $64$, $128$ and $512$, $q_{\rm max}=51.2$ and $u_{\rm max}=$ $2$, $0.8$, $0.4$ and $0.1$ (see text around Eq.\eqref{eq:indat}). The grid has $N=2^{12}$ points, $\epsilon=2 \pi/N$} 
\label{fig:EiffelTower}
\end{figure}

\begin{figure}
\includegraphics[width=4.2cm]{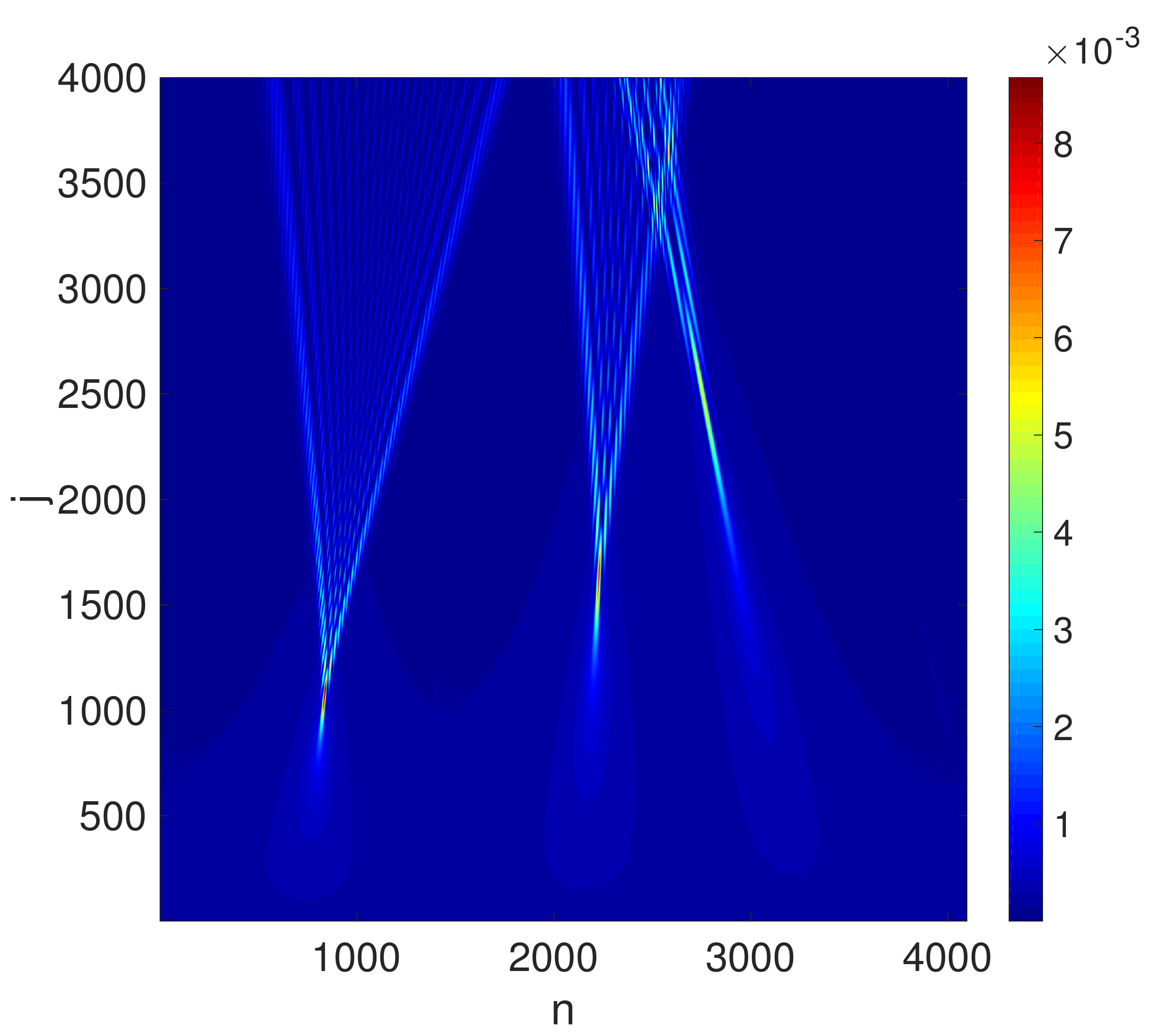}\hfill
\includegraphics[width=4.1cm]{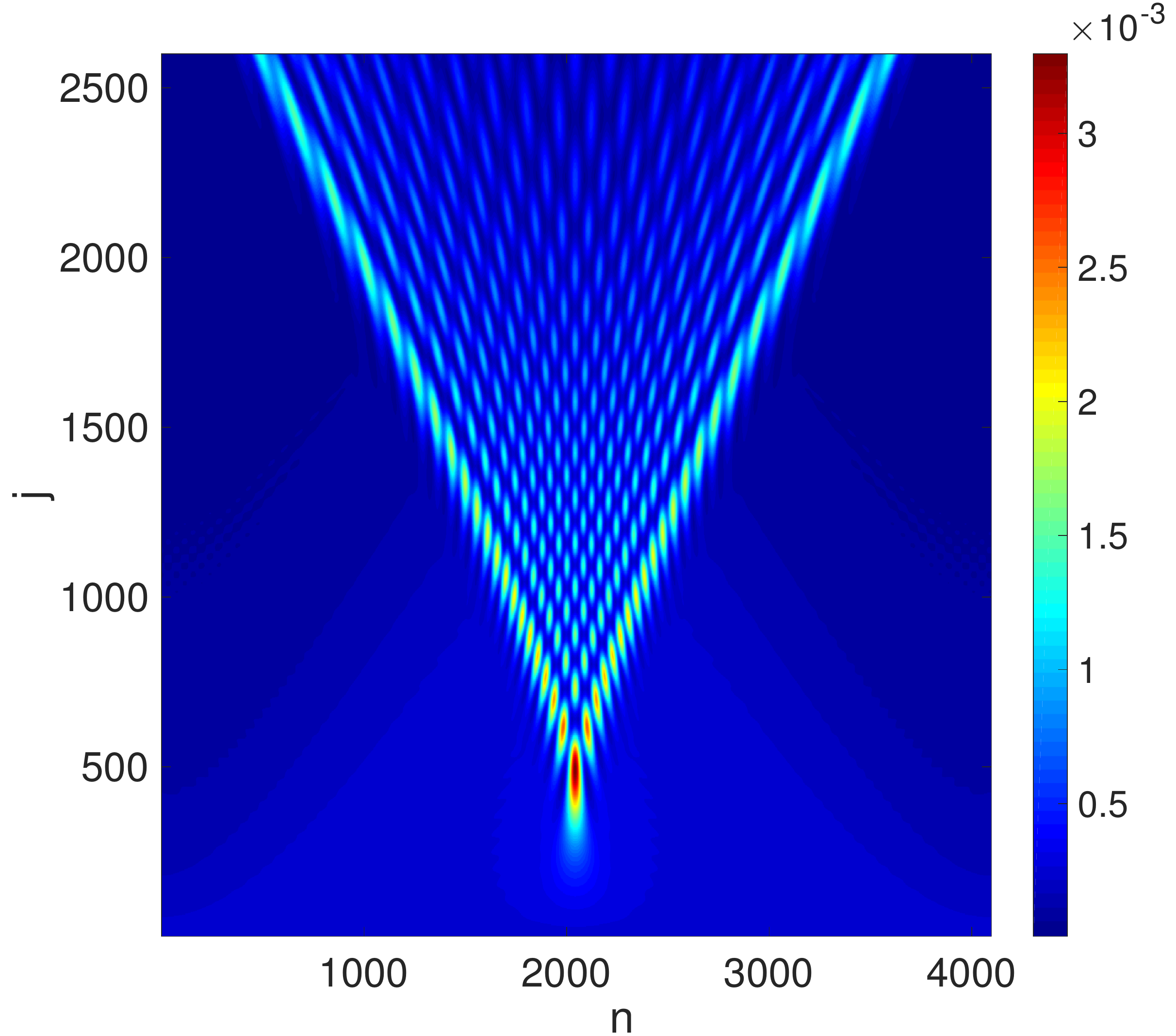}\hfill
\includegraphics[width=4.1cm]{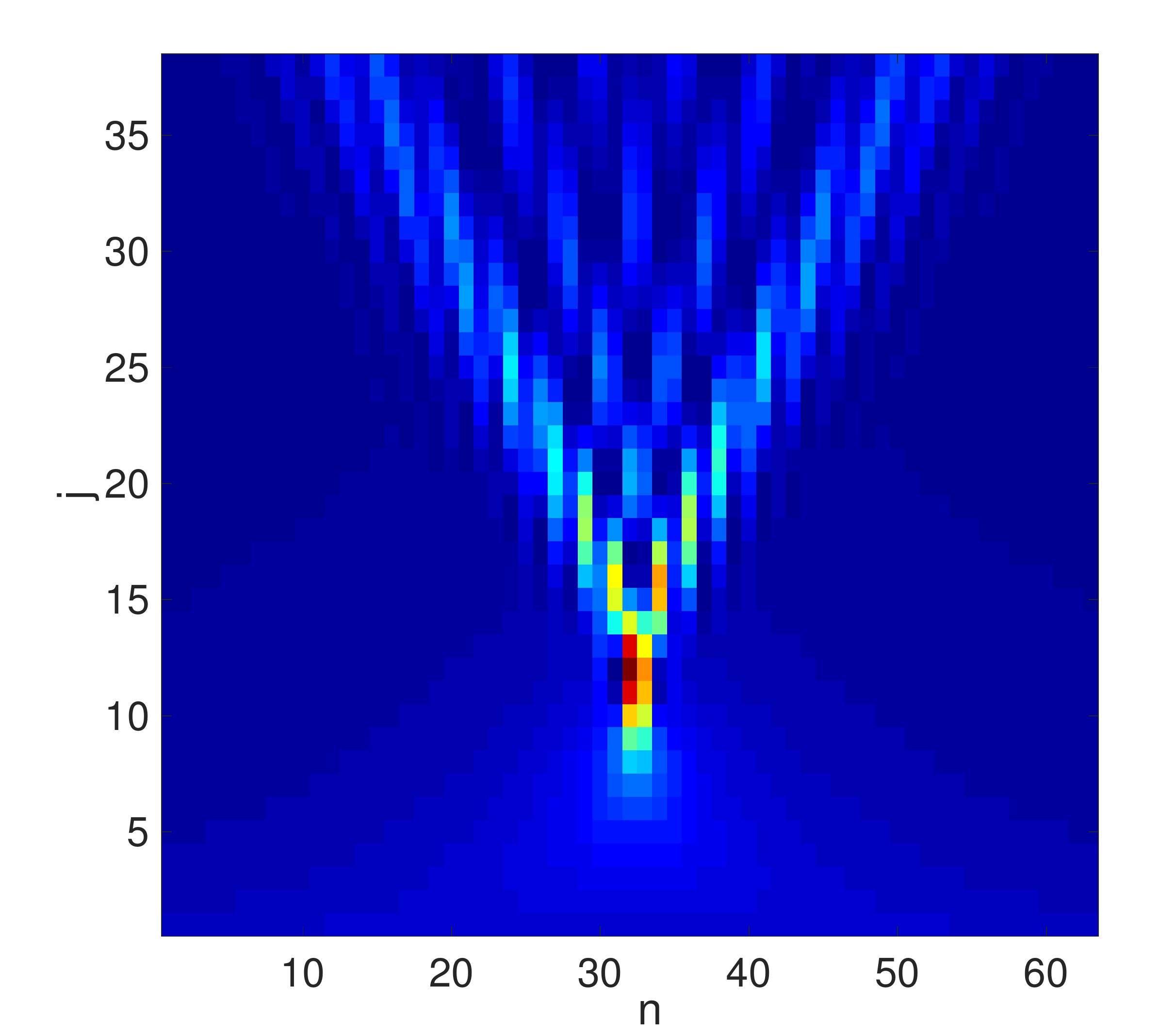}\hfill
\includegraphics[width=4.1cm]{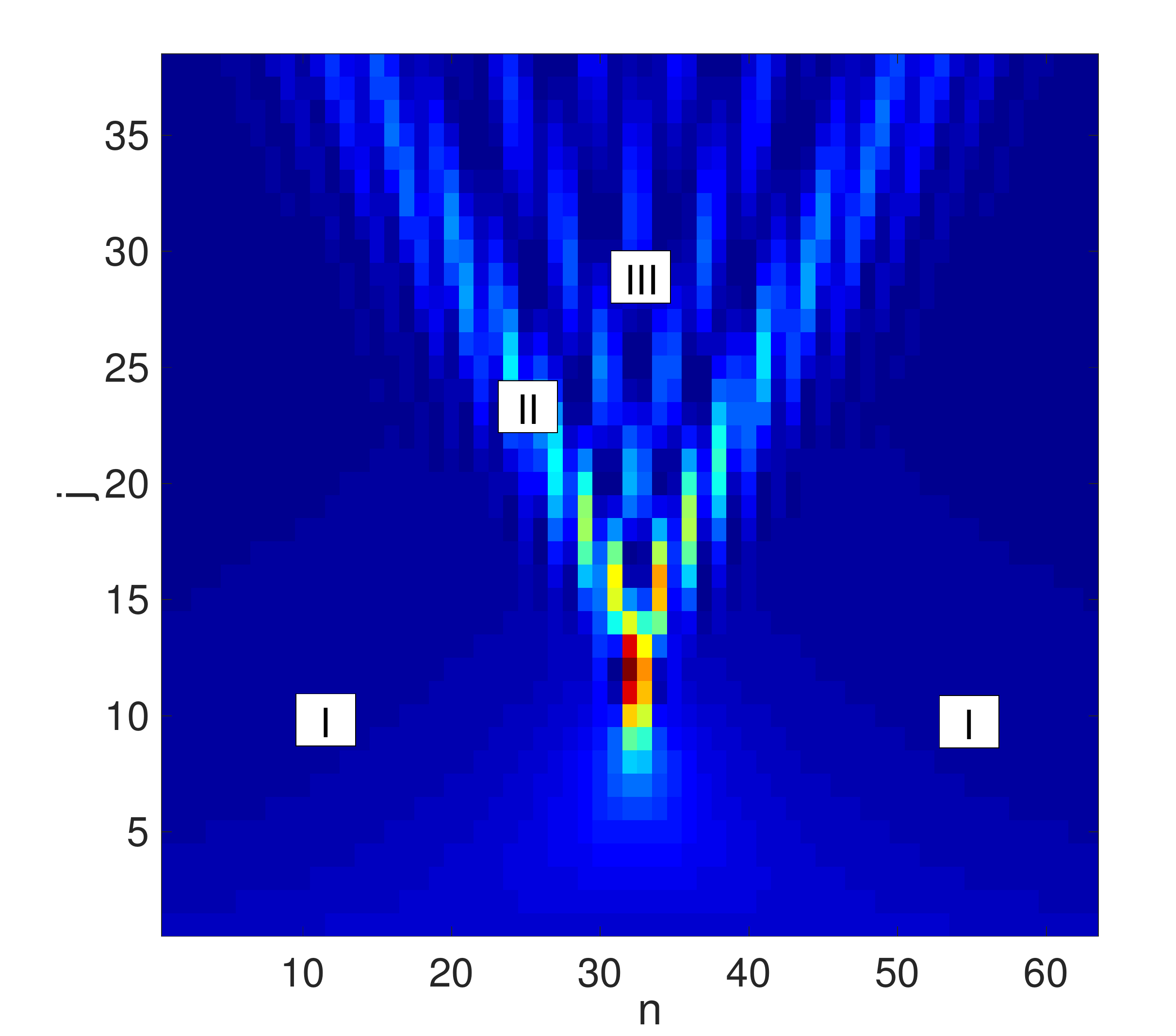}\hfill
\caption{(color online)  a) DTQW  with initial data ${\cos (x)+\frac{1}{3} \cos (3 x)+\frac{1}{2} \cos (2 x+0.9)}$ (same as Fig.\ref{fig:EiffelTower} but $u_{\rm max}=0.2$) b) Same as (a), but initial data: ${\cos (x)}$; c) Approximation by Pearcey's integral in  ($x,t$) space with $\varepsilon=0.05$ and d) zones of validity of approximations (see text)}
\label{fig:EiffelAndSimpleShock}
\end{figure}

\begin{figure}
\includegraphics[width=9.2cm]{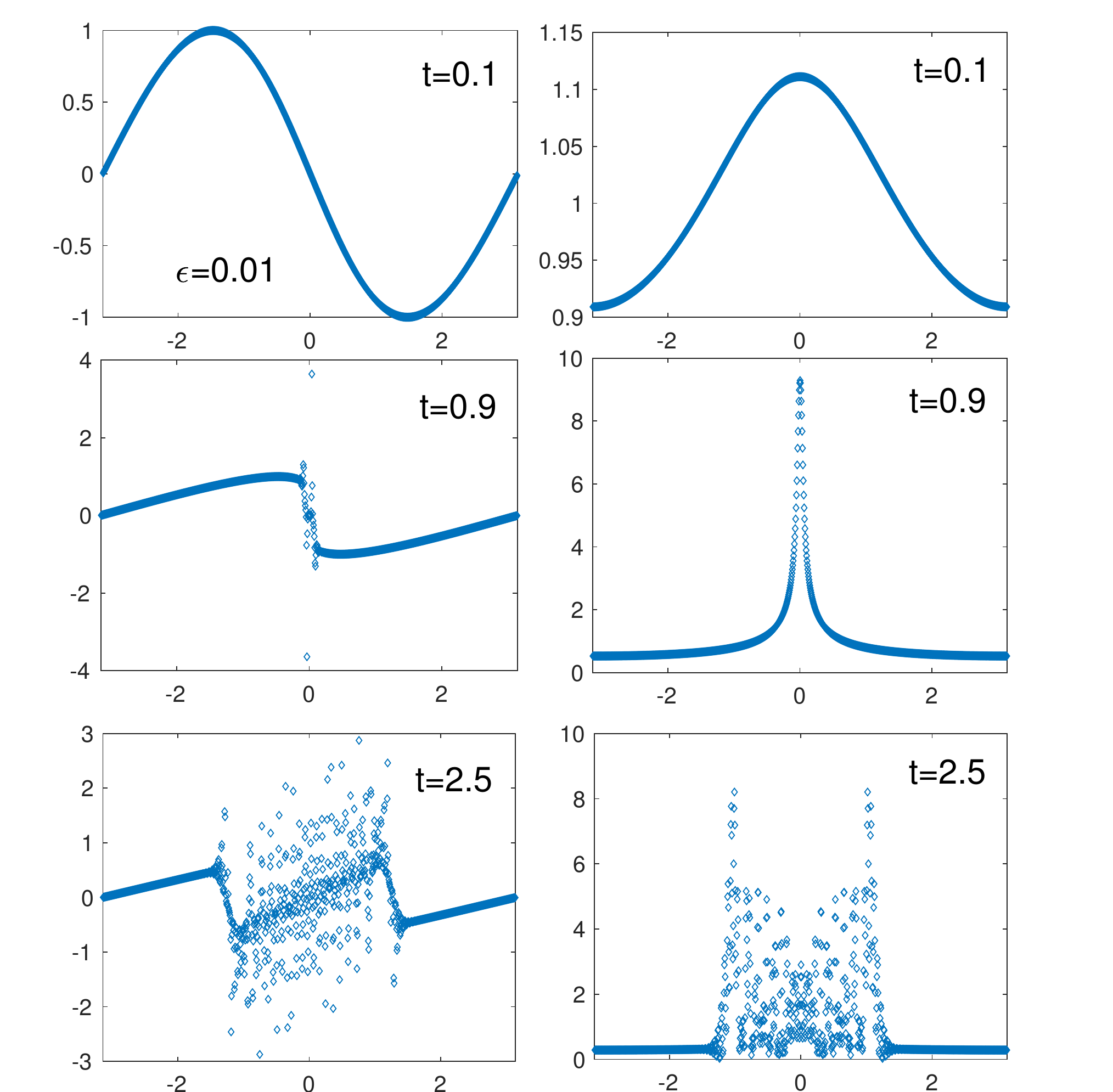}
\caption{ (color online)  Evolution of velocity $v = m^{-1} \partial_x \varphi$ (left) and density $n$ (right) for 3 values of time and $m^{-1}=0.01$ obtained from a numerical solution of Schr\"odinger equation $i \partial_t \psi=-\frac{1}{2 m}\partial_{xx}\psi$, with $\psi=n^{1/2}\exp{i \varphi}$ and initial data $\varphi=m \cos x$ and $n=1$.}
\label{fig:chocmadrhov}
\end{figure}

These figures show that the DTQW can indeed be used to simulate hydrodynamical shocks in a quantum fluid \cite{Hoefer2006,Wenjie2007}. 

\paragraph{Analytical shock computation in the Galilean regime}
We now present an analytical computation which reproduces the shock solution in the Galilean limit 
where the DTQW becomes a continuous time quantum walk and the Dirac equation goes, as shown in the supplemental material, into the 
($\hbar=0$) Schr\"odinger equation $i \partial_t \psi=-\frac{1}{2 m}\partial_{xx}\psi$.

The Green function for the Schr\"odinger equation reads
\begin{equation}
G_{0}(x,t|x_{0},t_{0})=\sqrt{\frac{m}{2i\pi(t-t_{0})}}e^{\frac{im(x-x_{0})^{2}}{2(t-t_{0})}}. \label{eq:evsch}
\end{equation}
The single-shock solution ($t_0 = 0$ and $u_{\rm max}=1$) thus reads
\begin{equation}
\psi ({\bf x},t) = \int_{-\infty }^{\infty } dy\,  \sqrt{\frac{m}{2i\pi t}} e^{i m\left(\frac{(y-x)^2}{2 t}+\cos (y)\right)} .
\end{equation}
In the large-$m$ limit, this integral can be computed by making use of 
methods that are standard in optics \cite{Berry1996}
and involve
Pearcey's integral \cite{Pearcey1946} defined by
\begin{equation}
I_{{\mathcal{P}}}(T,X)=\int_{-\infty }^{\infty } dy\,e^{i \left(Xy+T y^2+y^4\right)}\label{pearcey}.
\end{equation}
To wit, we set in the large-$m$ limit, $\psi ({\bf x},t)\approx A({\bf x},t) I_{{\mathcal{P}}}(-T (t),X ({\bf x},t))$
with $T(t)=a^{-\, \frac{1}{2}}\left(\frac{t-1}{2\varepsilon t}\right)$, $X({\bf x},t)=-a^{-\, \frac{1}{4}}\left(\frac{x}{\varepsilon t}\right)$
and 
$A({\bf x},t)={e^{i \left(1+\frac{x^2}{2 t}\right)/ \varepsilon }}(2 i \pi t \varepsilon \sqrt{a})^{-1/2}$
where $a=m/4!$ and $\varepsilon=1/m$.
In this way, Pearcey's integral Eq.\eqref{pearcey} alone can correctly reproduces the structure of the shock (see Figs.\ref{fig:EiffelAndSimpleShock}.c). 

Useful asymptotic expansions of $I_{{\mathcal{P}}}$
are given in \cite{Kaminski1989,Lopez2016}.
In particular, the steepest descent method can be directly used in zone $I$ (see fig.\ref{fig:EiffelAndSimpleShock}.d)
where $m \gg t/x^2$.
It yields the the following asymptotic form:
\begin{equation}
\psi_{I}({\bf x},t)\approx A({\bf x},t) \sqrt{\frac{-2i\pi}{\Phi''(u_{c})}}e^{i\Phi(u_{c})} 
\end{equation}
where $\Phi (u) =u^{4}-Tu^{2}+Xu$ and the single saddle-point $u_{c}$ obeys $\Phi'(u_{c})=0$.
Near the caustic, in zone $II$ of fig.\ref{fig:EiffelAndSimpleShock}.d, $2$ new saddle-points appear and
the wavefunction can be written in terms of the Airy function $Ai(x)=\frac{1}{\pi}\int_{-\infty }^{\infty }$ dt cos($\frac{t^{3}}{3}+xt$).
Well inside the caustic in zone $III$, the function can be written as the sum of $3$ interfering contributions 
(see Figs.\ref{fig:EiffelAndSimpleShock}.c and \ref{fig:EiffelAndSimpleShock} .d).

Details of the evolution of the density $n=|\psi|^2$ and velocity $v=\partial_x \phi/m$ of the Schr\"odinger shock are presented in Fig.\ref{fig:chocmadrhov}

\paragraph{Conclusion}
We have shown through a novel generalization of the Madelung transform that one of the simplest DTQWs on the line can be considered as a minimalist model of quantum fluids. This conclusion has been supported by numerical 
simulations which display the DTQW evolving an initial condition already considered in the literature into a quantum hydrodynamic shock  \cite{Hoefer2006}.\cite{Wenjie2007}.  We have also computed the asymptotic shock structure analytically in the non-relativistic limit and proposed an extensive discussion of this limit in the SM. 

Quantum walks have already been linked to with hydrodynamics in \cite{Succi2015} and \cite{Mezzacapo2015}, but these earlier results address the quantum Boltzmann equation and transport phenomena, and are thus quite different from those presented in this Letter. 

The present work should be extended to higher dimensions, higher spins and non-linear DTQWs \cite {DMDB15a} (or DTQWs with site to site interactions). One should also incorporate in the Madelung transform the natural coupling of DTQWs to gauge fields \cite{DMD14,ADelectromag16,ADMBD16a,AD17a}, thus obtaining novel models of superconducting quantum fluids or of quantum fluids in relativistic gravitational fields. 

\bibliographystyle{apsrev4-1}
\bibliography{Biblio.bib}

\begin{thebibliography}{33}%
\makeatletter
\providecommand \@ifxundefined [1]{%
 \@ifx{#1\undefined}
}%
\providecommand \@ifnum [1]{%
 \ifnum #1\expandafter \@firstoftwo
 \else \expandafter \@secondoftwo
 \fi
}%
\providecommand \@ifx [1]{%
 \ifx #1\expandafter \@firstoftwo
 \else \expandafter \@secondoftwo
 \fi
}%
\providecommand \natexlab [1]{#1}%
\providecommand \enquote  [1]{``#1''}%
\providecommand \bibnamefont  [1]{#1}%
\providecommand \bibfnamefont [1]{#1}%
\providecommand \citenamefont [1]{#1}%
\providecommand \href@noop [0]{\@secondoftwo}%
\providecommand \href [0]{\begingroup \@sanitize@url \@href}%
\providecommand \@href[1]{\@@startlink{#1}\@@href}%
\providecommand \@@href[1]{\endgroup#1\@@endlink}%
\providecommand \@sanitize@url [0]{\catcode `\\12\catcode `\$12\catcode
  `\&12\catcode `\#12\catcode `\^12\catcode `\_12\catcode `\%12\relax}%
\providecommand \@@startlink[1]{}%
\providecommand \@@endlink[0]{}%
\providecommand \url  [0]{\begingroup\@sanitize@url \@url }%
\providecommand \@url [1]{\endgroup\@href {#1}{\urlprefix }}%
\providecommand \urlprefix  [0]{URL }%
\providecommand \Eprint [0]{\href }%
\providecommand \doibase [0]{http://dx.doi.org/}%
\providecommand \selectlanguage [0]{\@gobble}%
\providecommand \bibinfo  [0]{\@secondoftwo}%
\providecommand \bibfield  [0]{\@secondoftwo}%
\providecommand \translation [1]{[#1]}%
\providecommand \BibitemOpen [0]{}%
\providecommand \bibitemStop [0]{}%
\providecommand \bibitemNoStop [0]{.\EOS\space}%
\providecommand \EOS [0]{\spacefactor3000\relax}%
\providecommand \BibitemShut  [1]{\csname bibitem#1\endcsname}%
\let\auto@bib@innerbib\@empty
\bibitem [{\citenamefont {Feynman}\ and\ \citenamefont
  {Hibbs.}(1965)}]{FeynHibbs65a}%
  \BibitemOpen
  \bibfield  {author} {\bibinfo {author} {\bibfnamefont {R.}~\bibnamefont
  {Feynman}}\ and\ \bibinfo {author} {\bibfnamefont {A.}~\bibnamefont
  {Hibbs.}},\ }\href@noop {} {\bibfield  {journal} {\bibinfo  {journal}
  {International Series in Pure and Applied Physics. McGraw-Hill Book Company}\
  } (\bibinfo {year} {1965})}\BibitemShut {NoStop}%
\bibitem [{\citenamefont {Aharonov}\ \emph {et~al.}(1993)\citenamefont
  {Aharonov}, \citenamefont {Davidovich},\ and\ \citenamefont
  {Zagury}}]{ADZ93a}%
  \BibitemOpen
  \bibfield  {author} {\bibinfo {author} {\bibfnamefont {Y.}~\bibnamefont
  {Aharonov}}, \bibinfo {author} {\bibfnamefont {L.}~\bibnamefont
  {Davidovich}}, \ and\ \bibinfo {author} {\bibfnamefont {N.}~\bibnamefont
  {Zagury}},\ }\href {\doibase 10.1103/PhysRevA.48.1687} {\bibfield  {journal}
  {\bibinfo  {journal} {Phys. Rev. A}\ }\textbf {\bibinfo {volume} {48}},\
  \bibinfo {pages} {1687} (\bibinfo {year} {1993})}\BibitemShut {NoStop}%
\bibitem [{\citenamefont {Meyer}(1996)}]{Meyer96a}%
  \BibitemOpen
  \bibfield  {author} {\bibinfo {author} {\bibfnamefont {D.~A.}\ \bibnamefont
  {Meyer}},\ }\href {\doibase 10.1007/BF02199356} {\bibfield  {journal}
  {\bibinfo  {journal} {Journal of Statistical Physics}\ }\textbf {\bibinfo
  {volume} {85}},\ \bibinfo {pages} {551} (\bibinfo {year} {1996})}\BibitemShut
  {NoStop}%
\bibitem [{\citenamefont {Schmitz}\ \emph {et~al.}(2009)\citenamefont
  {Schmitz}, \citenamefont {Matjeschk}, \citenamefont {Schneider},
  \citenamefont {Glueckert}, \citenamefont {Enderlein}, \citenamefont {Huber},\
  and\ \citenamefont {Schaetz}}]{Schmitz09a}%
  \BibitemOpen
  \bibfield  {author} {\bibinfo {author} {\bibfnamefont {H.}~\bibnamefont
  {Schmitz}}, \bibinfo {author} {\bibfnamefont {R.}~\bibnamefont {Matjeschk}},
  \bibinfo {author} {\bibfnamefont {C.}~\bibnamefont {Schneider}}, \bibinfo
  {author} {\bibfnamefont {J.}~\bibnamefont {Glueckert}}, \bibinfo {author}
  {\bibfnamefont {M.}~\bibnamefont {Enderlein}}, \bibinfo {author}
  {\bibfnamefont {T.}~\bibnamefont {Huber}}, \ and\ \bibinfo {author}
  {\bibfnamefont {T.}~\bibnamefont {Schaetz}},\ }\href {\doibase
  10.1103/PhysRevLett.103.090504} {\bibfield  {journal} {\bibinfo  {journal}
  {Phys. Rev. Lett.}\ }\textbf {\bibinfo {volume} {103}},\ \bibinfo {pages}
  {090504} (\bibinfo {year} {2009})}\BibitemShut {NoStop}%
\bibitem [{\citenamefont {{Z\"a}hringer}\ \emph {et~al.}(2010)\citenamefont
  {{Z\"a}hringer}, \citenamefont {Kirchmair}, \citenamefont {Gerritsma},
  \citenamefont {Solano}, \citenamefont {Blatt},\ and\ \citenamefont
  {Roos.}}]{Zahring10a}%
  \BibitemOpen
  \bibfield  {author} {\bibinfo {author} {\bibfnamefont {F.}~\bibnamefont
  {{Z\"a}hringer}}, \bibinfo {author} {\bibfnamefont {G.}~\bibnamefont
  {Kirchmair}}, \bibinfo {author} {\bibfnamefont {R.}~\bibnamefont
  {Gerritsma}}, \bibinfo {author} {\bibfnamefont {E.}~\bibnamefont {Solano}},
  \bibinfo {author} {\bibfnamefont {R.}~\bibnamefont {Blatt}}, \ and\ \bibinfo
  {author} {\bibfnamefont {C.}~\bibnamefont {Roos.}},\ }\href@noop {}
  {\bibfield  {journal} {\bibinfo  {journal} {Phys. Rev. Lett.}\ }\textbf
  {\bibinfo {volume} {104}},\ \bibinfo {pages} {100503} (\bibinfo {year}
  {2010})}\BibitemShut {NoStop}%
\bibitem [{\citenamefont {Schreiber}\ \emph {et~al.}(2010)\citenamefont
  {Schreiber}, \citenamefont {Cassemiro}, \citenamefont
  {Poto\ifmmode~\check{c}\else \v{c}\fi{}ek}, \citenamefont {G\'abris},
  \citenamefont {Mosley}, \citenamefont {Andersson}, \citenamefont {Jex},\ and\
  \citenamefont {Silberhorn}}]{Schreiber10a}%
  \BibitemOpen
  \bibfield  {author} {\bibinfo {author} {\bibfnamefont {A.}~\bibnamefont
  {Schreiber}}, \bibinfo {author} {\bibfnamefont {K.~N.}\ \bibnamefont
  {Cassemiro}}, \bibinfo {author} {\bibfnamefont {V.}~\bibnamefont
  {Poto\ifmmode~\check{c}\else \v{c}\fi{}ek}}, \bibinfo {author} {\bibfnamefont
  {A.}~\bibnamefont {G\'abris}}, \bibinfo {author} {\bibfnamefont {P.~J.}\
  \bibnamefont {Mosley}}, \bibinfo {author} {\bibfnamefont {E.}~\bibnamefont
  {Andersson}}, \bibinfo {author} {\bibfnamefont {I.}~\bibnamefont {Jex}}, \
  and\ \bibinfo {author} {\bibfnamefont {C.}~\bibnamefont {Silberhorn}},\
  }\href {\doibase 10.1103/PhysRevLett.104.050502} {\bibfield  {journal}
  {\bibinfo  {journal} {Phys. Rev. Lett.}\ }\textbf {\bibinfo {volume} {104}},\
  \bibinfo {pages} {050502} (\bibinfo {year} {2010})}\BibitemShut {NoStop}%
\bibitem [{\citenamefont {Karski}\ \emph {et~al.}(2009)\citenamefont {Karski},
  \citenamefont {F{\"o}rster}, \citenamefont {Choi}, \citenamefont {Steffen},
  \citenamefont {Alt}, \citenamefont {Meschede},\ and\ \citenamefont
  {Widera}}]{Karski09a}%
  \BibitemOpen
  \bibfield  {author} {\bibinfo {author} {\bibfnamefont {M.}~\bibnamefont
  {Karski}}, \bibinfo {author} {\bibfnamefont {L.}~\bibnamefont {F{\"o}rster}},
  \bibinfo {author} {\bibfnamefont {J.-M.}\ \bibnamefont {Choi}}, \bibinfo
  {author} {\bibfnamefont {A.}~\bibnamefont {Steffen}}, \bibinfo {author}
  {\bibfnamefont {W.}~\bibnamefont {Alt}}, \bibinfo {author} {\bibfnamefont
  {D.}~\bibnamefont {Meschede}}, \ and\ \bibinfo {author} {\bibfnamefont
  {A.}~\bibnamefont {Widera}},\ }\href {\doibase 10.1126/science.1174436}
  {\bibfield  {journal} {\bibinfo  {journal} {Science}\ }\textbf {\bibinfo
  {volume} {325}},\ \bibinfo {pages} {174} (\bibinfo {year} {2009})},\ \Eprint
  {http://arxiv.org/abs/http://science.sciencemag.org/content/325/5937/174.full.pdf}
  {http://science.sciencemag.org/content/325/5937/174.full.pdf} \BibitemShut
  {NoStop}%
\bibitem [{\citenamefont {Sansoni}\ \emph {et~al.}(2012)\citenamefont
  {Sansoni}, \citenamefont {Sciarrino}, \citenamefont {Vallone}, \citenamefont
  {Mataloni}, \citenamefont {Crespi}, \citenamefont {Ramponi},\ and\
  \citenamefont {Osellame}}]{Sansoni11a}%
  \BibitemOpen
  \bibfield  {author} {\bibinfo {author} {\bibfnamefont {L.}~\bibnamefont
  {Sansoni}}, \bibinfo {author} {\bibfnamefont {F.}~\bibnamefont {Sciarrino}},
  \bibinfo {author} {\bibfnamefont {G.}~\bibnamefont {Vallone}}, \bibinfo
  {author} {\bibfnamefont {P.}~\bibnamefont {Mataloni}}, \bibinfo {author}
  {\bibfnamefont {A.}~\bibnamefont {Crespi}}, \bibinfo {author} {\bibfnamefont
  {R.}~\bibnamefont {Ramponi}}, \ and\ \bibinfo {author} {\bibfnamefont
  {R.}~\bibnamefont {Osellame}},\ }\href {\doibase
  10.1103/PhysRevLett.108.010502} {\bibfield  {journal} {\bibinfo  {journal}
  {Phys. Rev. Lett.}\ }\textbf {\bibinfo {volume} {108}},\ \bibinfo {pages}
  {010502} (\bibinfo {year} {2012})}\BibitemShut {NoStop}%
\bibitem [{\citenamefont {Sanders}\ \emph {et~al.}(2003)\citenamefont
  {Sanders}, \citenamefont {Bartlett}, \citenamefont {Tregenna},\ and\
  \citenamefont {Knight.}}]{Sanders03a}%
  \BibitemOpen
  \bibfield  {author} {\bibinfo {author} {\bibfnamefont {B.}~\bibnamefont
  {Sanders}}, \bibinfo {author} {\bibfnamefont {S.}~\bibnamefont {Bartlett}},
  \bibinfo {author} {\bibfnamefont {B.}~\bibnamefont {Tregenna}}, \ and\
  \bibinfo {author} {\bibfnamefont {P.}~\bibnamefont {Knight.}},\ }\href@noop
  {} {\bibfield  {journal} {\bibinfo  {journal} {Phys. {R}ev. A}\ }\textbf
  {\bibinfo {volume} {67}},\ \bibinfo {pages} {042305} (\bibinfo {year}
  {2003})}\BibitemShut {NoStop}%
\bibitem [{\citenamefont {Perets}\ \emph {et~al.}(2008)\citenamefont {Perets},
  \citenamefont {Lahini}, \citenamefont {Pozzi}, \citenamefont {Sorel},
  \citenamefont {Morandotti},\ and\ \citenamefont {Silberberg}}]{Perets08a}%
  \BibitemOpen
  \bibfield  {author} {\bibinfo {author} {\bibfnamefont {B.}~\bibnamefont
  {Perets}}, \bibinfo {author} {\bibfnamefont {Y.}~\bibnamefont {Lahini}},
  \bibinfo {author} {\bibfnamefont {F.}~\bibnamefont {Pozzi}}, \bibinfo
  {author} {\bibfnamefont {M.}~\bibnamefont {Sorel}}, \bibinfo {author}
  {\bibfnamefont {R.}~\bibnamefont {Morandotti}}, \ and\ \bibinfo {author}
  {\bibfnamefont {Y.}~\bibnamefont {Silberberg}},\ }\href@noop {} {\bibfield
  {journal} {\bibinfo  {journal} {Phys. Rev. Lett.}\ }\textbf {\bibinfo
  {volume} {100}},\ \bibinfo {pages} {170506} (\bibinfo {year}
  {2008})}\BibitemShut {NoStop}%
\bibitem [{\citenamefont {Ambainis.}(2007)}]{Amb07a}%
  \BibitemOpen
  \bibfield  {author} {\bibinfo {author} {\bibfnamefont {A.}~\bibnamefont
  {Ambainis.}},\ }\href@noop {} {\bibfield  {journal} {\bibinfo  {journal}
  {SIAM Journal on Computing}\ }\textbf {\bibinfo {volume} {37}},\ \bibinfo
  {pages} {210} (\bibinfo {year} {2007})}\BibitemShut {NoStop}%
\bibitem [{\citenamefont {Magniez}\ \emph {et~al.}( ACM)\citenamefont
  {Magniez}, \citenamefont {A.~Nayak},\ and\ \citenamefont {Santha}}]{MNRS07a}%
  \BibitemOpen
  \bibfield  {author} {\bibinfo {author} {\bibfnamefont {F.}~\bibnamefont
  {Magniez}}, \bibinfo {author} {\bibfnamefont {J.~R.}\ \bibnamefont
  {A.~Nayak}}, \ and\ \bibinfo {author} {\bibfnamefont {M.}~\bibnamefont
  {Santha}},\ }\href@noop {} {\bibfield  {journal} {\bibinfo  {journal} {SIAM
  Journal on Computing - Proceedings of the thirty-ninth annual ACM symposium
  on Theory of computing}\ } (\bibinfo {year} {New {Y}ork, 2007.
  ACM.})}\BibitemShut {NoStop}%
\bibitem [{\citenamefont {Aslangul.}(2005)}]{Aslangul05a}%
  \BibitemOpen
  \bibfield  {author} {\bibinfo {author} {\bibfnamefont {C.}~\bibnamefont
  {Aslangul.}},\ }\href@noop {} {\bibfield  {journal} {\bibinfo  {journal}
  {Journal of Physics A: Mathematical and Theoretical}\ }\textbf {\bibinfo
  {volume} {38}},\ \bibinfo {pages} {1} (\bibinfo {year} {2005})}\BibitemShut
  {NoStop}%
\bibitem [{\citenamefont {Bose.}(2003)}]{Bose03a}%
  \BibitemOpen
  \bibfield  {author} {\bibinfo {author} {\bibfnamefont {S.}~\bibnamefont
  {Bose.}},\ }\href@noop {} {\bibfield  {journal} {\bibinfo  {journal} {Phys.
  Rev. Lett.}\ }\textbf {\bibinfo {volume} {91}},\ \bibinfo {pages} {207901}
  (\bibinfo {year} {2003})}\BibitemShut {NoStop}%
\bibitem [{\citenamefont {Burgarth}(2006)}]{Burg06a}%
  \BibitemOpen
  \bibfield  {author} {\bibinfo {author} {\bibfnamefont {D.}~\bibnamefont
  {Burgarth}},\ }\href@noop {} {\bibfield  {journal} {\bibinfo  {journal}
  {University College London}\ }\textbf {\bibinfo {volume} {PhD thesis}}
  (\bibinfo {year} {2006})}\BibitemShut {NoStop}%
\bibitem [{\citenamefont {Bose}(2007)}]{Bose07a}%
  \BibitemOpen
  \bibfield  {author} {\bibinfo {author} {\bibfnamefont {S.}~\bibnamefont
  {Bose}},\ }\href {\doibase 10.1080/00107510701342313} {\bibfield  {journal}
  {\bibinfo  {journal} {Contemporary Physics}\ }\textbf {\bibinfo {volume}
  {48}},\ \bibinfo {pages} {13} (\bibinfo {year} {2007})},\ \Eprint
  {http://arxiv.org/abs/http://dx.doi.org/10.1080/00107510701342313}
  {http://dx.doi.org/10.1080/00107510701342313} \BibitemShut {NoStop}%
\bibitem [{\citenamefont {Collini}\ \emph {et~al.}(2010)\citenamefont
  {Collini}, \citenamefont {Wong}, \citenamefont {Wilk}, \citenamefont {Curmi},
  \citenamefont {Brumer},\ and\ \citenamefont {Scholes}}]{Collini10a}%
  \BibitemOpen
  \bibfield  {author} {\bibinfo {author} {\bibfnamefont {E.}~\bibnamefont
  {Collini}}, \bibinfo {author} {\bibfnamefont {C.}~\bibnamefont {Wong}},
  \bibinfo {author} {\bibfnamefont {K.}~\bibnamefont {Wilk}}, \bibinfo {author}
  {\bibfnamefont {P.}~\bibnamefont {Curmi}}, \bibinfo {author} {\bibfnamefont
  {P.}~\bibnamefont {Brumer}}, \ and\ \bibinfo {author} {\bibfnamefont
  {G.}~\bibnamefont {Scholes}},\ }\href@noop {} {\bibfield  {journal} {\bibinfo
   {journal} {Nature}\ ,\ \bibinfo {pages} {644}} (\bibinfo {year}
  {2010})}\BibitemShut {NoStop}%
\bibitem [{\citenamefont {Engel}\ \emph {et~al.}(2007)\citenamefont {Engel},
  \citenamefont {Calhoun}, \citenamefont {Read}, \citenamefont {Ahn},
  \citenamefont {Mancal}, \citenamefont {Cheng}, \citenamefont {Blankenship},\
  and\ \citenamefont {Fleming}}]{Engel07a}%
  \BibitemOpen
  \bibfield  {author} {\bibinfo {author} {\bibfnamefont {G.~S.}\ \bibnamefont
  {Engel}}, \bibinfo {author} {\bibfnamefont {T.~R.}\ \bibnamefont {Calhoun}},
  \bibinfo {author} {\bibfnamefont {E.~L.}\ \bibnamefont {Read}}, \bibinfo
  {author} {\bibfnamefont {T.-K.}\ \bibnamefont {Ahn}}, \bibinfo {author}
  {\bibfnamefont {T.}~\bibnamefont {Mancal}}, \bibinfo {author} {\bibfnamefont
  {Y.-C.}\ \bibnamefont {Cheng}}, \bibinfo {author} {\bibfnamefont {R.~E.}\
  \bibnamefont {Blankenship}}, \ and\ \bibinfo {author} {\bibfnamefont {G.~R.}\
  \bibnamefont {Fleming}},\ }\href {http://dx.doi.org/10.1038/nature05678}
  {\bibfield  {journal} {\bibinfo  {journal} {Nature}\ }\textbf {\bibinfo
  {volume} {446}},\ \bibinfo {pages} {782} (\bibinfo {year}
  {2007})}\BibitemShut {NoStop}%
\bibitem [{\citenamefont {Molfetta}\ \emph {et~al.}(2014)\citenamefont
  {Molfetta}, \citenamefont {Brachet},\ and\ \citenamefont {Debbasch}}]{DMD14}%
  \BibitemOpen
  \bibfield  {author} {\bibinfo {author} {\bibfnamefont {G.~D.}\ \bibnamefont
  {Molfetta}}, \bibinfo {author} {\bibfnamefont {M.}~\bibnamefont {Brachet}}, \
  and\ \bibinfo {author} {\bibfnamefont {F.}~\bibnamefont {Debbasch}},\ }\href
  {\doibase http://dx.doi.org/10.1016/j.physa.2013.11.036} {\bibfield
  {journal} {\bibinfo  {journal} {Physica A: Statistical Mechanics and its
  Applications}\ }\textbf {\bibinfo {volume} {397}},\ \bibinfo {pages} {157 }
  (\bibinfo {year} {2014})}\BibitemShut {NoStop}%
\bibitem [{\citenamefont {Coles}\ and\ \citenamefont
  {Spencer}(2003)}]{Coles2003}%
  \BibitemOpen
  \bibfield  {author} {\bibinfo {author} {\bibfnamefont {P.}~\bibnamefont
  {Coles}}\ and\ \bibinfo {author} {\bibfnamefont {K.}~\bibnamefont
  {Spencer}},\ }\href {\doibase 10.1046/j.1365-8711.2003.06529.x} {\bibfield
  {journal} {\bibinfo  {journal} {Monthly Notices of the Royal Astronomical
  Society}\ }\textbf {\bibinfo {volume} {342}},\ \bibinfo {pages} {176}
  (\bibinfo {year} {2003})},\ \Eprint
  {http://arxiv.org/abs/http://mnras.oxfordjournals.org/content/342/1/176.full.pdf+html}
  {http://mnras.oxfordjournals.org/content/342/1/176.full.pdf+html}
  \BibitemShut {NoStop}%
\bibitem [{\citenamefont {Sikivie}\ and\ \citenamefont
  {Yang}(2009)}]{Sikivie2009}%
  \BibitemOpen
  \bibfield  {author} {\bibinfo {author} {\bibfnamefont {P.}~\bibnamefont
  {Sikivie}}\ and\ \bibinfo {author} {\bibfnamefont {Q.}~\bibnamefont {Yang}},\
  }\href {\doibase 10.1103/PhysRevLett.103.111301} {\bibfield  {journal}
  {\bibinfo  {journal} {Phys. Rev. Lett.}\ }\textbf {\bibinfo {volume} {103}},\
  \bibinfo {pages} {111301} (\bibinfo {year} {2009})}\BibitemShut {NoStop}%
\bibitem [{\citenamefont {Hoefer}\ \emph {et~al.}(2006)\citenamefont {Hoefer},
  \citenamefont {Ablowitz}, \citenamefont {Coddington}, \citenamefont
  {Cornell}, \citenamefont {Engels},\ and\ \citenamefont
  {Schweikhard}}]{Hoefer2006}%
  \BibitemOpen
  \bibfield  {author} {\bibinfo {author} {\bibfnamefont {M.~A.}\ \bibnamefont
  {Hoefer}}, \bibinfo {author} {\bibfnamefont {M.~J.}\ \bibnamefont
  {Ablowitz}}, \bibinfo {author} {\bibfnamefont {I.}~\bibnamefont
  {Coddington}}, \bibinfo {author} {\bibfnamefont {E.~A.}\ \bibnamefont
  {Cornell}}, \bibinfo {author} {\bibfnamefont {P.}~\bibnamefont {Engels}}, \
  and\ \bibinfo {author} {\bibfnamefont {V.}~\bibnamefont {Schweikhard}},\
  }\href {\doibase 10.1103/PhysRevA.74.023623} {\bibfield  {journal} {\bibinfo
  {journal} {Phys. Rev. A}\ }\textbf {\bibinfo {volume} {74}},\ \bibinfo
  {pages} {023623} (\bibinfo {year} {2006})}\BibitemShut {NoStop}%
\bibitem [{\citenamefont {Wan}\ \emph {et~al.}(2007)\citenamefont {Wan},
  \citenamefont {Jia},\ and\ \citenamefont {Fleischer}}]{Wenjie2007}%
  \BibitemOpen
  \bibfield  {author} {\bibinfo {author} {\bibfnamefont {W.}~\bibnamefont
  {Wan}}, \bibinfo {author} {\bibfnamefont {S.}~\bibnamefont {Jia}}, \ and\
  \bibinfo {author} {\bibfnamefont {J.~W.}\ \bibnamefont {Fleischer}},\ }\href
  {http://dx.doi.org/10.1038/nphys486} {\bibfield  {journal} {\bibinfo
  {journal} {Nat Phys}\ }\textbf {\bibinfo {volume} {3}},\ \bibinfo {pages}
  {46} (\bibinfo {year} {2007})}\BibitemShut {NoStop}%
\bibitem [{\citenamefont {Berry}\ and\ \citenamefont
  {Klein}(1996)}]{Berry1996}%
  \BibitemOpen
  \bibfield  {author} {\bibinfo {author} {\bibfnamefont {M.~V.}\ \bibnamefont
  {Berry}}\ and\ \bibinfo {author} {\bibfnamefont {S.}~\bibnamefont {Klein}},\
  }\href {http://www.pnas.org/content/93/6/2614.abstract} {\bibfield  {journal}
  {\bibinfo  {journal} {Proceedings of the National Academy of Sciences}\
  }\textbf {\bibinfo {volume} {93}},\ \bibinfo {pages} {2614} (\bibinfo {year}
  {1996})},\ \Eprint
  {http://arxiv.org/abs/http://www.pnas.org/content/93/6/2614.full.pdf}
  {http://www.pnas.org/content/93/6/2614.full.pdf} \BibitemShut {NoStop}%
\bibitem [{\citenamefont {Pearcey}(1946)}]{Pearcey1946}%
  \BibitemOpen
  \bibfield  {author} {\bibinfo {author} {\bibfnamefont {T.}~\bibnamefont
  {Pearcey}},\ }\href {\doibase 10.1080/14786444608561335} {\bibfield
  {journal} {\bibinfo  {journal} {The London, Edinburgh, and Dublin
  Philosophical Magazine and Journal of Science}\ }\textbf {\bibinfo {volume}
  {37}},\ \bibinfo {pages} {311} (\bibinfo {year} {1946})},\ \Eprint
  {http://arxiv.org/abs/http://dx.doi.org/10.1080/14786444608561335}
  {http://dx.doi.org/10.1080/14786444608561335} \BibitemShut {NoStop}%
\bibitem [{\citenamefont {Kaminski}(1989)}]{Kaminski1989}%
  \BibitemOpen
  \bibfield  {author} {\bibinfo {author} {\bibfnamefont {D.}~\bibnamefont
  {Kaminski}},\ }\href {\doibase 10.1137/0520066} {\bibfield  {journal}
  {\bibinfo  {journal} {SIAM Journal on Mathematical Analysis}\ }\textbf
  {\bibinfo {volume} {20}},\ \bibinfo {pages} {987} (\bibinfo {year} {1989})},\
  \Eprint {http://arxiv.org/abs/http://dx.doi.org/10.1137/0520066}
  {http://dx.doi.org/10.1137/0520066} \BibitemShut {NoStop}%
\bibitem [{\citenamefont {L{\'o}pez}\ and\ \citenamefont
  {Pagola}(2016)}]{Lopez2016}%
  \BibitemOpen
  \bibfield  {author} {\bibinfo {author} {\bibfnamefont {J.~L.}\ \bibnamefont
  {L{\'o}pez}}\ and\ \bibinfo {author} {\bibfnamefont {P.~J.}\ \bibnamefont
  {Pagola}},\ }\href {\doibase http://dx.doi.org/10.1016/j.amc.2015.11.080}
  {\bibfield  {journal} {\bibinfo  {journal} {Applied Mathematics and
  Computation}\ }\textbf {\bibinfo {volume} {275}},\ \bibinfo {pages} {404 }
  (\bibinfo {year} {2016})}\BibitemShut {NoStop}%
\bibitem [{\citenamefont {Succi}\ \emph {et~al.}(2015)\citenamefont {Succi},
  \citenamefont {Fillion-Gourdeau},\ and\ \citenamefont
  {Palpacelli}}]{Succi2015}%
  \BibitemOpen
  \bibfield  {author} {\bibinfo {author} {\bibfnamefont {S.}~\bibnamefont
  {Succi}}, \bibinfo {author} {\bibfnamefont {F.}~\bibnamefont
  {Fillion-Gourdeau}}, \ and\ \bibinfo {author} {\bibfnamefont
  {S.}~\bibnamefont {Palpacelli}},\ }\href {\doibase
  10.1140/epjqt/s40507-015-0025-1} {\bibfield  {journal} {\bibinfo  {journal}
  {EPJ Quantum Technology}\ }\textbf {\bibinfo {volume} {2}},\ \bibinfo {pages}
  {1} (\bibinfo {year} {2015})}\BibitemShut {NoStop}%
\bibitem [{\citenamefont {Mezzacapo}\ \emph {et~al.}(2015)\citenamefont
  {Mezzacapo}, \citenamefont {Sanz}, \citenamefont {Lamata}, \citenamefont
  {Egusquiza}, \citenamefont {Succi},\ and\ \citenamefont
  {Solano}}]{Mezzacapo2015}%
  \BibitemOpen
  \bibfield  {author} {\bibinfo {author} {\bibfnamefont {A.}~\bibnamefont
  {Mezzacapo}}, \bibinfo {author} {\bibfnamefont {M.}~\bibnamefont {Sanz}},
  \bibinfo {author} {\bibfnamefont {L.}~\bibnamefont {Lamata}}, \bibinfo
  {author} {\bibfnamefont {I.~L.}\ \bibnamefont {Egusquiza}}, \bibinfo {author}
  {\bibfnamefont {S.}~\bibnamefont {Succi}}, \ and\ \bibinfo {author}
  {\bibfnamefont {E.}~\bibnamefont {Solano}},\ }\href {\doibase
  10.1038/srep13153} {\bibfield  {journal} {\bibinfo  {journal} {Scientific
  Reports}\ }\textbf {\bibinfo {volume} {5}},\ \bibinfo {pages} {13153}
  (\bibinfo {year} {2015})}\BibitemShut {NoStop}%
\bibitem [{\citenamefont {Di~Molfetta}\ \emph {et~al.}(2015)\citenamefont
  {Di~Molfetta}, \citenamefont {Debbasch},\ and\ \citenamefont
  {Brachet}}]{DMDB15a}%
  \BibitemOpen
  \bibfield  {author} {\bibinfo {author} {\bibfnamefont {G.}~\bibnamefont
  {Di~Molfetta}}, \bibinfo {author} {\bibfnamefont {F.}~\bibnamefont
  {Debbasch}}, \ and\ \bibinfo {author} {\bibfnamefont {M.}~\bibnamefont
  {Brachet}},\ }\href {\doibase 10.1103/PhysRevE.92.042923} {\bibfield
  {journal} {\bibinfo  {journal} {Phys. Rev. E}\ }\textbf {\bibinfo {volume}
  {92}},\ \bibinfo {pages} {042923} (\bibinfo {year} {2015})}\BibitemShut
  {NoStop}%
\bibitem [{\citenamefont {Arnault}\ and\ \citenamefont
  {Debbasch}(2016)}]{ADelectromag16}%
  \BibitemOpen
  \bibfield  {author} {\bibinfo {author} {\bibfnamefont {P.}~\bibnamefont
  {Arnault}}\ and\ \bibinfo {author} {\bibfnamefont {F.}~\bibnamefont
  {Debbasch}},\ }\href {\doibase 10.1103/PhysRevA.93.052301} {\bibfield
  {journal} {\bibinfo  {journal} {Phys. Rev. A}\ }\textbf {\bibinfo {volume}
  {93}},\ \bibinfo {pages} {052301} (\bibinfo {year} {2016})}\BibitemShut
  {NoStop}%
\bibitem [{\citenamefont {Arnault}\ \emph {et~al.}(2016)\citenamefont
  {Arnault}, \citenamefont {Di~Molfetta}, \citenamefont {Brachet},\ and\
  \citenamefont {Debbasch}}]{ADMBD16a}%
  \BibitemOpen
  \bibfield  {author} {\bibinfo {author} {\bibfnamefont {P.}~\bibnamefont
  {Arnault}}, \bibinfo {author} {\bibfnamefont {G.}~\bibnamefont
  {Di~Molfetta}}, \bibinfo {author} {\bibfnamefont {M.}~\bibnamefont
  {Brachet}}, \ and\ \bibinfo {author} {\bibfnamefont {F.}~\bibnamefont
  {Debbasch}},\ }\href {\doibase 10.1103/PhysRevA.94.012335} {\bibfield
  {journal} {\bibinfo  {journal} {Phys. Rev. A}\ }\textbf {\bibinfo {volume}
  {94}},\ \bibinfo {pages} {012335} (\bibinfo {year} {2016})}\BibitemShut
  {NoStop}%
\bibitem [{\citenamefont {Arnault}\ and\ \citenamefont
  {Debbasch}(2017)}]{AD17a}%
  \BibitemOpen
  \bibfield  {author} {\bibinfo {author} {\bibfnamefont {P.}~\bibnamefont
  {Arnault}}\ and\ \bibinfo {author} {\bibfnamefont {F.}~\bibnamefont
  {Debbasch}},\ }\href {\doibase http://dx.doi.org/10.1016/j.aop.2017.04.003}
  {\bibfield  {journal} {\bibinfo  {journal} {Annals of Physics}\ }\textbf
  {\bibinfo {volume} {383}},\ \bibinfo {pages} {645 } (\bibinfo {year}
  {2017})}\BibitemShut {NoStop}%
\end{thebibliography}%

\end{document}


\title{Supplementary material to "Quantum walk hydrodynamics"}

\author{Mohamed Hatifi}
\affiliation{Aix Marseille Universit\'e, Institut Fresnel CNRS UMR 7249, 13013 Marseille, France}
\author{Giuseppe Di Molfetta}
\affiliation{Laboratoire d'Informatique Fondamentale, CaNa Research Group, 
UMR7279 CNRS and Universit\'e Aix-Marseille,
163 Avenue de Luminy, 13288 Marseille France}

\author{Fabrice Debbasch}
\affiliation{LERMA, UMR 8112, UPMC and Observatoire de Paris, 61 Avenue de l'Observatoire  75014 Paris, France}
\author{Marc Brachet} 
\affiliation{Laboratoire de Physique Statistique, 
\'{E}cole Normale Sup{\'e}rieure, 
PSL Research University;
UPMC Univ Paris 06, Sorbonne Universit\'{e}s;
Universit\'{e} Paris Diderot, Sorbonne Paris-Cit\'{e};
CNRS;
24 Rue Lhomond, 75005 Paris, France
}

\begin{abstract}
This SM provides supplementary information on "Non relativistic limit" of the Quantum Walks Hydrodynamics. 
\end{abstract}

\maketitle

\section{Non relativistic limit}

\subsection{Wave equation}

Dirac equation $i\gamma^{\mu}\partial_{\mu} \psi-m\psi=0$ 
reads, in component terms and in units where $\hbar = 1$ but $c \ne 1$:

\begin{eqnarray}
\frac{1}{c} \partial_t \Psi_{L} - \partial_x \Psi_{L} & = & -imc \Psi_{R} \nonumber \\
\frac{1}{c} \partial_t \Psi_{R} + \partial_x \Psi_{R} & = & -imc \Psi_{L} 
\end{eqnarray}

Each component obeys the same Klein-Gordon (KG) equation
\begin{equation}
\frac{1}{c^2} \partial_{tt} \Psi_{L/R} - \partial_{xx} \Psi_{L/R} = - m^2 c^2 \Psi_{L/R}.
\end{equation}

The non relativistic limit is obtained as follows. We write $\Psi_{L/R} = {\bar \Psi}_{L/R} \exp(-i mc^2 t)$. From now on, all scalings below characterize the action of differential operators on ${\bar \Psi}_{L/R}$. We suppose 
\begin{equation}
\frac{1}{mc^2} \partial_t = \mathcal{O}(\nu^2)
\end{equation}
where $\nu$ is an infinitesimal. This scaling traces that the energy of $\Psi$ is close to the rest mass energy. The difference will be the kinetic energy 
which, for a free particle, scales as the squared momentum $p^2$. Thus, $p$ must scale as $\nu$.  

Injecting he above scaling in the KG equation for ${\bar \Psi}_{L/R}$ shows that ${\bar \Psi}_{L/R}$ obey the Schr\"odinger equation when $\nu$ goes to zero.
We now also compute for future use the lowest order terms in the difference ${\bar \Psi}_{L} - {\bar \Psi}_{R}$.
The Dirac equation can be rewritten as
\begin{eqnarray}
{\bar \Psi}_{R} & = & {\bar \Psi}_{L} + \frac{1}{i m c} \partial_x {\bar \Psi}_{L} + \frac{i}{mc^2} \partial_t {\bar \Psi}_{L}
\nonumber \\ 
{\bar \Psi}_{L} & = & {\bar \Psi}_{R} - \frac{1}{i m c} \partial_x {\bar \Psi}_{R} + \frac{i}{mc^2} \partial_t {\bar \Psi}_{R}\label{eq:5}.
\end{eqnarray}

In the right-hand sides of the previous equations, the second terms are of order $\nu$ while the third terms are of order $\nu^2$. 
This is so because $- i \hbar \partial_x$ is the impulse operator and $p = mc + \mathcal{O}(\nu)$ (see remark above). 
Using the Schr{\"o}dinger equation makes the orders of the terms more explicit:

\begin{eqnarray}
{\bar \Psi}_{R} & = & {\bar \Psi}_{L} + \frac{1}{i m c} \partial_x {\bar \Psi}_{L} - \frac{1}{2 m^2 c^2} \partial_{xx} {\bar \Psi}_{L} + \mathcal{O}(\nu^3) \nonumber \\ 
{\bar \Psi}_{L} & = & {\bar \Psi}_{R} -  \frac{1}{i m c} \partial_x {\bar \Psi}_{R} - \frac{1}{2 m^2 c^2} \partial_{xx} {\bar \Psi}_{R} + \mathcal{O}(\nu^3)\label{eq:9}.
\label{eq:psimeqpsip}
\end{eqnarray}

\subsection{Lagrangian density}

Let us start the discussion by keeping only the terms of order $\nu$ in (\ref{eq:psimeqpsip}). The two wave-function components are equal at order 0 in $\nu$ and thus, at this order, have the same moduli and phases. We want to compute the differences between the moduli and the differences between the phases at first order in $\nu$. This is best done in the following way. Write ${\bar \Psi}_{L} = r \exp(i \phi)$ and ${\bar \Psi}_{R} = (r + \delta r) \exp\left(i (\phi + \delta \phi)\right)$. 
Thus (see eq. \eqref{eq:psi} and definitions just above) 
$r^2=\rho_L$, $(r +\delta r)^2 = \rho_R$, $\phi=\varphi_L -m c^2 t$ and $\delta \phi=\varphi_-$.

Inserting this into (\ref{eq:psimeqpsip}) and keeping only first order terms leads to: 
\begin{equation}
r \exp (i \phi) \left(i \delta \phi + \frac{\delta r}{r} \right) = \frac{1}{i m c} \partial_x {\bar \Psi}_{L},
 \end{equation}
from which one gets:
\begin{equation}
\delta \phi  = - \frac{1}{2 m c r^2} \left(  {\bar \Psi}_{L}^{*} \partial_x {\bar \Psi}_{L} + \partial_x  {\bar \Psi}_{L}^{*} {\bar \Psi}_{L}\right).
 \end{equation}
The difference $\delta r$ can be obtained in the same manner:
 \begin{equation}
\frac{\delta r}{r}  = - \frac{1}{2 i m c r^2} \left( {\bar \Psi}_{L}^{*} \partial_x {\bar \Psi}_{L} - \partial_x  {\bar \Psi}_{L}^{*} {\bar \Psi}_{L}\right).
 \end{equation}
 
This transcribes into:
 \begin{equation}
\delta \phi  = - \frac{1}{mc}\, \frac{1}{r}\, \frac{\partial r}{\partial x} 
 \end{equation}
 and
\begin{equation}
\frac{\delta r}{r}  =  \frac{1}{mc}\,\frac{\partial \phi}{\partial x} .
 \end{equation} 

It is straightforward (but tedious) to compute in the same manner the differences in moduli and phases at second order in $\nu$. One finds:

 \begin{equation}
\delta \phi  = - \frac{1}{mc}\, \frac{1}{r}\, \frac{\partial r}{\partial x} 
 - \frac{1}{2m^2c^2}\, \frac{\partial^2\phi}{\partial x^2}
 \end{equation}
 and
\begin{eqnarray}
\frac{\delta r}{r}  & =  & \frac{1}{mc}\,\frac{\partial \phi}{\partial x} + \frac{1}{2 mc^2}\, \left(\frac{\partial \phi}{\partial x}\right)^2 \nonumber \\
& + & 
\frac{1}{2m^2c^2 r^2}\, \left(  \left(\frac{\partial r}{\partial x}\right)^2
- r \, \frac{\partial^2 r}{\partial x^2}
\right)
 \end{eqnarray} 

Inserting this  into the Dirac Lagrangian in hydrodynamical form $\mathcal{L}$ and making use of  the equations of motion 
(4)-(5) of the main text 
delivers the expression of the Lagrangian density at second order in $\nu$:
$\mathcal{L_S}={\mathcal{L}}/{4}=-({r^2(\partial_x\phi)^2}+{(\partial_{x}r)^2})/{2 m}- r^2\partial_t\phi$
that, using Madelung's relation $\psi=r \exp(i \phi)$, coincides
with the usual ($\hbar=1$) Schr\"odinger Lagrangian:
$\mathcal{L_S}= {i}(\overline{\psi}{\partial_t \psi}-\psi \partial_t \overline{\psi})/2-|\partial_x\psi|^{2}/2m$.
Note that the equations of motion need to be used to pass from the Dirac Lagrangian to the Schr\"odinger Lagrangian because
the two spinor components of the Dirac wave function degenerate into a single Schr\"odinger wave-function.

\subsection{Hydrodynamical variables}
At second order in $\nu$ the hydrodynamical variables defined in the main text read:
\begin{equation}
n=2 r^2 + \frac{2r^2}{mc}\, \frac{\partial \phi}{\partial x}
+  \frac{1}{m^2 c^2}\, \left[r^2 \left(\frac{\partial \phi}{\partial x}\right)^2 +\left(\frac{\partial r}{\partial x}\right)^2 - r \frac{\partial^2 r}{\partial x^2}\right]
   \label{eq:n}
 \end{equation}

\begin{equation}
u^0=1+ \frac{1}{2 m^2 c^2}\, \left(\frac{\partial \phi}{\partial x}\right)^2
\label{eq:u0}
 \end{equation}
 
 \begin{equation}
u^1= \frac{1}{mc} \frac{\partial \phi}{\partial x}+ \frac{1}{2 m^2 c^2 r^2}\,  \left[\left(\frac{\partial r}{\partial x}\right)^2- r \frac{\partial^2 r}{\partial x^2} \right]
\label{eq:u1}
 \end{equation}
 
 \begin{equation}
w=  2 m c^2 r^2+2 c  r^2 \frac{\partial \phi}{\partial x}+
\frac{r^2 (\frac{\partial \phi}{\partial x})^2  -  r \frac{\partial^2 r}{\partial x^2}}{m}
\label{eq:w}
 \end{equation}

 \bibliographystyle{apsrev4-1}
\bibliography{library}